\documentclass[superscriptaddress,nofootinbib,twocolumn,prb,letterpaper]{revtex4-1}

\usepackage{amsmath,amssymb}
\usepackage{graphicx}
\usepackage[usenames, dvipsnames]{xcolor}
\usepackage{hyperref}
\usepackage{url}
\usepackage[utf8]{inputenc}
\usepackage{slashed}
\usepackage{subfigure}
\usepackage{slashed,bbm}
\usepackage{graphics,psfrag,epsfig}
\usepackage{dsfont}
\usepackage{setspace}
\usepackage{mathtools}

\graphicspath{{./}{./plots/}}

\newcommand{\ONa}{\mathrm{O}(N_1)}
\newcommand{\ONb}{\mathrm{O}(N_2)}
\newcommand{\ONab}{\mathrm{O}(N_1 + N_2)}
\newcommand{\ON}{\mathrm{O}(N)}
\newcommand{\Ofive}{\mathrm{O}(5)}

\newcommand{\Nf}{N_\mathrm{f}}

\begin{document}

\title{Compatible orders and fermion-induced emergent symmetry in Dirac systems}

%
\author{Lukas Janssen}
\affiliation{Institut f\"ur Theoretische Physik, Technische Universit\"at Dresden, 01062 Dresden, Germany}
\author{Igor~F.~Herbut}
\affiliation{Department of Physics, Simon Fraser University, Burnaby, British Columbia, Canada V5A 1S6}
\author{Michael~M.~Scherer}
\affiliation{Institut f\"ur Theoretische Physik, Universit\"at K\"oln, 50937 K\"oln, Germany}
%

\begin{abstract}
We study the quantum multicritical point in a (2+1)D Dirac system between the semimetallic phase and two ordered phases that are characterized by anticommuting mass terms with $\mathrm{O}(N_1)$ and $\mathrm{O}(N_2)$ symmetry, respectively.
Using $\epsilon$ expansion around the upper critical space-time dimension of four, we demonstrate the existence of a stable renormalization-group fixed point, enabling a direct and continuous transition between the two ordered phases directly at the multicritical point.
This point is found to be characterized by an emergent $\mathrm{O}(N_1+N_2)$ symmetry for arbitrary values of $N_1$ and $N_2$ and fermion flavor numbers $N_\mathrm{f}$, as long as the corresponding representation of the Clifford algebra exists.
Small $\mathrm{O}(N)$-breaking perturbations near the chiral $\mathrm{O}(N)$ fixed point are therefore irrelevant.
This result can be traced back to the presence of gapless Dirac degrees of freedom at criticality, and it is in clear contrast to the purely bosonic $\mathrm{O}(N)$ fixed point, which is stable only when $N < 3$.
As a by-product, we obtain predictions for the critical behavior of the chiral $\mathrm{O}(N)$ universality classes for arbitrary $N$ and fermion flavor number $N_\mathrm{f}$.
Implications for critical Weyl and Dirac systems in (3+1)D are also briefly discussed.
\end{abstract}

\maketitle

\paragraph*{Introduction.}

The interplay and competition of different ordering tendencies in many-body systems are the source of various exciting phenomena, including unconventional superconductivity, the nature of quantum spin liquids, and the physics of deconfined criticality. These notoriously challenging problems sometimes become theoretically accessible when an emergent higher symmetry can be found. The complex phase diagram of the high-$T_\mathrm{c}$ superconductors, for instance, has been argued to be deducible from an emergent symmetry in which the $\mathrm{O}(3)$ N\'eel and $\mathrm{U}(1)$ superconducting order parameters are combined into a five-tuplet which turns out to be a vector under $\mathrm{O}(5)$.\cite{Zhang1997}
Numerical simulations of the deconfined critical point between the N\'eel and valence bond solid orders on the square lattice also find evidence for an emergent $\mathrm{O}(5)$ symmetry.\cite{Nahum:2015vka} The emergence of this symmetry can be made natural by postulating duality relations between the bosonic gauge theory describing the deconfined critical point and certain fermionic theories.\cite{Wang:2017txt, Janssen:2017eeu}
Similarly, recent quantum Monte Carlo simulations of Dirac fermions in $2+1$ dimensions find a direct and continuous transition between $\mathrm{O}(3)$ and $\mathds{Z}_2$ ordered phases with an emergent $\mathrm{O}(4)$ symmetry at criticality.\cite{Sato:2017tgx}

Despite the general interest, however, a simple model in which emergent $\mathrm{O}(N)$ symmetry with $N\geq 4$ can be explicitly shown, appears to be still lacking.
This is certainly true within the standard Landau-Ginzburg-Wilson approach, in which a continuous quantum phase transition is assumed to be described by bosonic order-parameter fluctuations alone:\cite{Pelissetto:2000ek,herbut2007,Eichhorn:2013zza} The purely bosonic $\ON$ fixed point in $2+1$ dimensions is unstable to small perturbations that break the $\ON$ symmetry for all $N>3$, and presumably even the Heisenberg fixed point is unstable to a cubic anisotropy.\cite{Carmona:1999rm,Calabrese:2002bm}
In this work, we demonstrate that the stability under symmetry-breaking perturbations significantly changes in the presence of gapless fermionic degrees of freedom. In particular, we demonstrate that the \emph{chiral} $\ON$ fixed point, in which the bosonic order parameter is coupled to $\Nf$ flavors of massless Dirac fermions in $2+1$ dimensions, is stable to
perturbations that break the $\ON$ symmetry.
This adds a prime example to the general observation of fermion-induced symmetry enhancement in quantum critical Dirac systems.\cite{li2015,Scherer:2016zwz, Ponte:2012ru, Grover:2013rc, Li:2017dkj}
Our results can be immediately applied to the triple point between the semimetallic and the $\mathrm O(3)$ 
N\'eel and $\mathrm U(1)$ 
Kekul\'e valence bond solid phases on the honeycomb lattice. The crucial ingredient here is the anticommuting nature of the corresponding Dirac mass terms, enabling us to combine them into a single order parameter that becomes a vector under $\Ofive$. The triple point is characterized by emergent $\Ofive$ symmetry and features a continuous and direct transition between the ordered phases, as long as the system is tuned directly through the multicritical point. The corresponding universal exponents define the chiral $\Ofive$ universality class for which we provide estimates. A similar reasoning applies to the multicritical point between the Dirac semimetal and the $\mathrm O(3)$ and $\mathds{Z}_2$ ordered phases.\cite{Sato:2017tgx}
%

\paragraph*{Anticommuting Dirac masses.}

Consider the gapless Dirac Hamiltonian in $D=d+1$ space-time dimensions
\begin{align}
	\mathcal H_0(\vec p) = \alpha_i p_i, \qquad i = 1,\dots,d.
\end{align}
We assume the summation convention over repeated indices.
The matrices $\alpha_i$ fulfill the Clifford algebra $\{\alpha_i, \alpha_j \} = 2 \delta_{ij} \mathds 1_{d_\gamma}$, with $d_\gamma$ being the dimension of the representation.
The Hamiltonian $\mathcal H_0$ can be gapped out by adding (explicit or dynamical) mass terms
\begin{align}
	\mathcal H_m = m_a \beta_a^\phi + m_b \beta_b^\chi, \  a = 1, ...\,, N_1,\ b = 1,...\,, N_2.
\end{align}
The mass operators $\beta_a^\phi$ and $\beta_b^\chi$ anticommute with $\mathcal H_0$
%
%
and among themselves.
%
%
Their commutators $M_{a a'}^\phi = \frac{i}{2} [\beta_a^\phi, \beta_{a'}^\phi]$ and $M_{b b'}^\chi = \frac{i}{2} [\beta_b^\chi, \beta_{b'}^\chi]$ commute with $\mathcal H_0$ and generate an $\ONa \oplus \ONb$ symmetry under which the mass operators $\beta_a^\phi$ and $\beta_b^\chi$ transform as vectors.
In this work, we assume the masses $m_a$ and $m_b$ to be \emph{compatible}, i.e., their mass operators $\beta_a^\phi$ and $\beta_b^\chi$ also mutually anticommute,
\begin{align} \label{eq:anticomm-a-b}
	\{\beta_a^\phi, \beta_b^\chi\} = 0, \qquad \text{for all $a$, $b$.}
\end{align}
The massive Hamiltonian $\mathcal H_0 + \mathcal H_m$ has the spectrum $\varepsilon_{\vec{p}} = \pm \sqrt{p^2 + m_a^2 + m_b^2}$, and $m_a$ and $m_b$ do not compete.\cite{ryu2009}

The Dirac system with $\ONa \oplus \ONb$ anticommuting mass terms in $d$ spatial dimensions requires $N_1 + N_2 + d$ anticommuting $d_\gamma \times d_\gamma$ matrices. The number of Dirac fermion components $d_\gamma$ is therefore\cite{das2014}
\begin{align} \label{eq:dgamma}
	d_\gamma \geq 2^{\lfloor \frac{N_1 + N_2 + d}{2}\rfloor},
\end{align}
where $\lfloor \, \cdot \, \rfloor$ denotes the floor function. In the situation relevant for graphene we have $d=2$ and $d_\gamma = 8$, and the maximal number of anticommuting mass terms is thus $N_1 + N_2 \leq 5$, which is consistent with the known classification of the 36 mass terms of spin-$1/2$ fermions on the honeycomb lattice.\cite{ryu2009}
We note that, in general, smaller (real) representations are possible if one employs a Nambu particle-hole construction, and the right-hand-side of Eq.~\eqref{eq:dgamma} is then to be replaced by $n/2$, where $n$ is the dimension of the irreducible real representation of the Clifford algebra $C(d,N_1+N_2)$.\cite{Herbut:2011th}

\paragraph*{Model.}

We study the system of gapless Dirac fermions $\Psi$ and $\Psi^\dagger$ coupled to compatible order parameters $\phi \equiv \sum_a^{N_1}\phi_a \beta_a^\phi$ and $\chi \equiv \sum_b^{N_2} \chi_b \beta^\chi_b$ with anticommuting mass operators $\beta_a^\phi$ and $\beta_b^\chi$. This is described by the Lagrangian
\begin{align}
	\mathcal L_\mathrm{F} = \Psi^\dagger \left[\partial_\tau + \mathcal H_0(-i \vec \nabla) +  g_1 \phi + g_2 \chi \right] \Psi \,,
\end{align}
with the Yukawa-type couplings $g_1$ and $g_2$ parametrizing the coupling to the fluctuating boson fields $\phi$ and $\chi$. When radiative corrections are taken into account, the latter receive their own dynamics  as well as bosonic selfinteractions. We therefore include already from the outset all symmetry-allowed terms that may become generated by the fluctuations, up to fourth order in the fields,
\begin{align}
	\mathcal L_\mathrm{B} & =
	\frac{1}{2} \phi_a \left(-\partial_\mu^2 + r_1 \right) \phi_a
	+ \frac{1}{2} \chi_b \left(-\partial_\mu^2 + r_2 \right) \chi_b
	\nonumber \\ &\quad
	+ \lambda_{1}\left(\phi_a^2 \right)^2
	+ \lambda_{2}\left(\chi_b^2\right)^2
	+ 2\lambda_{3} \, \phi_a^2 \chi_b^2
	\,,
\end{align}
with $(\partial_\mu) \equiv (\partial_\tau, \vec \nabla)$ and $\mu = 0,1,\dots,d$.

The action of the full system then is given by
$
	S = \int d\tau d^d \vec x \left(\mathcal L_F+\mathcal L_B \right).
$
This action describes a theory space with an explicit $\ONa \oplus \ONb$ symmetry generated by the $[N_1(N_1-1)+ N_2(N_2-1)]/2$ commutators $M_{aa'}^\phi$ and $M_{bb'}^\chi$, where $a < a' = 1,\dots, N_1$ and $b < b' = 1,\dots,N_2$, respectively.
It includes an $\ONab$-invariant subspace, which is achieved by choosing
%
$g_1 = g_2, \ \lambda_1 = \lambda_2 = \lambda_{3}, \ r_1 = r_2$.
%
The additional generators promoting $\ONa \oplus \ONb$ to $\ONab$ are the $N_1\cdot N_2$ operators $M_{ab}^{\phi\chi} = \frac{i}{2} [ \beta_a^\phi, \beta_b^\chi ]$. They rotate $\phi$ and $\chi$ into each other, allowing to construct a $(N_1 + N_2)$-tuplet $(\phi_a, \chi_b)$, which transforms as a vector under $\ONab$.

The Yukawa-type couplings and the bosonic selfinteractions have scaling dimension $[g_1^2] = [g_2^2] = [\lambda_1] = [\lambda_2] = [\lambda_{3}] = 3-d$, and therefore become simultaneously marginal in three spatial dimensions. In the following, we will mainly consider $1<d<3$, with a focus on $d=2$. In this case, all couplings become relevant in the renormalization group (RG) sense. However, the ratios $g_1^2/r_1$ and $g_2^2/r_2$, describing the effective strengths of the interaction,\cite{ZinnJustin:1991yn} are irrelevant for large enough $r_1$ and $r_2$, respectively. In this limit, the massive boson fields can be integrated out, leaving behind the stable noninteracting semimetal phase. The latter is trivially within the $\ONab$-invariant subspace.
Upon lowering $r_1$ ($r_2$) towards zero, on the other hand, the $\ONa$ [$\ONb$] part of the full symmetry group becomes spontaneously broken to a residual $\mathrm{O}(N_1-1)$ [$\mathrm{O}(N_2-1)$] symmetry, characterized by a finite vacuum expectation value $\langle \phi_a \rangle \neq 0$ [$\langle \chi_b \rangle \neq 0$] for some $a$ ($b$). Due to the Yukawa couplings $g_1$ and $g_2$, the condensation of the bosonic fields simultaneously opens a mass gap of the Dirac fermions. $r_1$ and $r_2$ are the tuning parameters for the corresponding quantum phase transitions. We are interested in the behavior of the fermionic \emph{multicritical} point, \cite{Roy:2011pg,Roy:2013aya,Classen:2015ssa,Classen:2015mar} in which both $r_1$ and $r_2$ are tuned to their critical values.

An important example in $d=2$, to which the above model can be immediately applied, is the physics of interacting spin-$1/2$ fermions on the honeycomb lattice. One of the 56 distinct five-tuplets of pairwise anticommuting mass operators consists of, for instance, the three components of the N\'eel order parameter and the two components of the Kekul\'e valence bond solid order parameter.\cite{ryu2009} In the notation of Ref.~\onlinecite{Herbut:2009vu}, these masses read
\begin{align}
	(\beta_a^\phi)_{a=1,2,3}= \vec \sigma \otimes \gamma_0,\
	(\beta_b^\chi)_{b=1,2} = \mathds 1_2 \otimes ( i \gamma_0 \gamma_3, i \gamma_0 \gamma_5)\,.\notag
\end{align}
In the lattice model of Ref.~\onlinecite{Sato:2017tgx}, the $\mathrm O(2)$ symmetry in the Kekul\'e sector is explicitly broken down to $\mathds Z_2 \simeq \mathrm O(1)$.\cite{footnote0}
Note that the chiral $\ON$ fixed point in the presence of a small $\ON$-breaking perturbation can be understood as a multicritical point and our model therefore applies also to this problem.

\paragraph*{Flow equations.}

The existence of a unique upper critical dimension allows a controlled expansion in powers of $\epsilon = 3 - d$.
Integrating over the momentum shell from $\Lambda/b$ to $\Lambda$, we obtain the flow equations to the leading order,
\begin{align}
	\dot{g}_1^2 & = (\epsilon - \eta_{1} - 2 \eta_\Psi ) g_1^2 + 2 (N_1 - 2) g_1^4 + 2 N_2 g_1^2 g_2^2 \,, \label{eq:g1} \displaybreak[0] \\
	\dot{g}_2^2 & = (\epsilon - \eta_{2} - 2 \eta_\Psi ) g_2^2 + 2 (N_2 - 2) g_2^4 + 2 N_1 g_1^2 g_2^2\,, \label{eq:g2}
\end{align}
where $\dot{g}_i^2 \equiv \frac{d g_i^2}{d \ln b}$, $i=1,2$, and accordingly
\begin{align}
	\dot{\lambda}_1&=(\epsilon -  2\eta_1) \lambda_1 - 4(N_1 + 8)\lambda_1^2 - 4N_2\lambda_{3}^2 + \Nf g_1^4\,, \label{eq:lambda1} \displaybreak[0] \\
	\dot{\lambda}_2 &=(\epsilon -  2\eta_2) \lambda_2 - 4(N_2 + 8)\lambda_2^2 - 4N_1\lambda_{3}^2 + \Nf g_2^4\,, \label{eq:lambda2}
\displaybreak[0]\\
	\dot{\lambda}_{3}& =(\epsilon - \eta_1 - \eta_2) \lambda_{3} - 16 \lambda_{3}^2 - 4(N_1 + 2)\lambda_1\lambda_{3}
	\nonumber\\ & \quad - 4(N_2 + 2)\lambda_2\lambda_{3} + \Nf g_1^2 g_2^2\,, \label{eq:lambda12}
\end{align}
with fermion anomalous dimension $\eta_\Psi = (N_1 g_1^2+N_2 g_2^2)/2$ and order-parameter anomalous dimensions $\eta_{i} = 2\Nf g_i^2$.
In order to arrive at the above equations, we have rescaled the couplings as $g_i^2 \Lambda^{\eta_i + 2 \eta_\Psi - \epsilon} /(8\pi^2) \mapsto g_i^2$ and $\lambda_i \Lambda^{2\eta_i - \epsilon} /(8\pi^2) \mapsto \lambda_i$ with $\eta_3 = (\eta_1+\eta_2)/2$,
and abbreviated $\Nf \coloneqq d_\gamma/4$. We have tuned both $\phi$ and $\chi$ to criticality by setting $r_1 = r_2 = 0$.
This system of flow equations simplifies to various known results within respective limits: For $g_1, g_2 \to 0$, we recover the flow equations of the purely bosonic system with $\ONa \oplus \ONb$ symmetry.\cite{Pelissetto:2000ek,Calabrese:2002bm,herbut2007} For $N_1 = 1$ and $N_2 = 2$, Eqs.~\eqref{eq:g1}--\eqref{eq:lambda2} also agree with the stability analysis of the $\mathds Z_2 \times \mathrm O(2)$ Gross-Neveu-Yukawa theory.\cite{Roy:2011pg, footnote1}
Finally, in the isotropic limit $g_1 = g_2$ and $\lambda_1 = \lambda_2 = \lambda_{3}$, the flow equations reduce to the known equations for the chiral Ising, chiral XY, and chiral Heisenberg universality classes for $N_1+N_2 = 1$, $2$, and $3$, respectively.\cite{Herbut:2009vu,Mihaila:2017ble,Zerf:2017zqi}

\paragraph*{Stability of the isotropic fixed point.}

\begin{figure}
\includegraphics[scale=0.87]{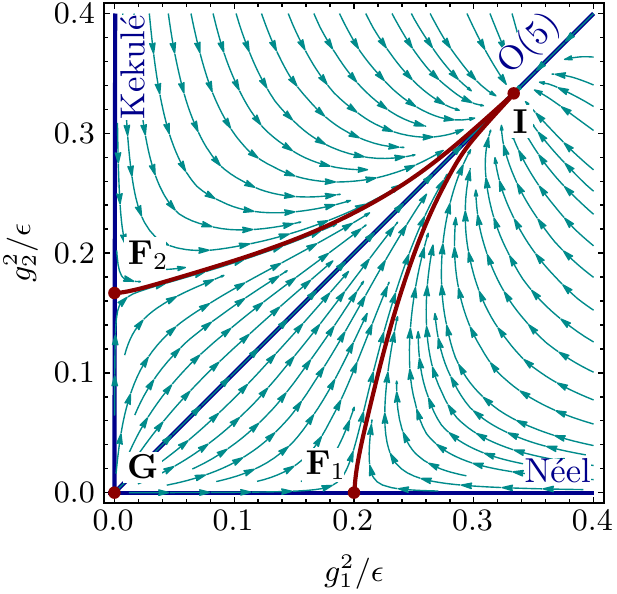}
\caption{RG flow diagram in $g_1^2$-$g_2^2$ sector near the multicritical point between $\mathrm{O}(3)$ N\'eel antiferromagnet and $\mathrm{O}(2)$ Kekul\'e order, using $d_\gamma = 4\Nf = 8$. The decoupled fixed points $\mathrm F_1$ and $\mathrm{F}_2$ describe the individual chiral Heisenberg and XY universality classes. The unique stable fixed point is the isotropic fixed point $\mathrm I$, characterized by an emergent $\mathrm{O}(5)$ symmetry.}
\label{fig:flow}
\end{figure}

As the bosonic selfinteractions do not feed back into the Yukawa-coupling sector, the latter can be solved independently. In addition to the noninteracting Gaussian fixed point at $g_1^2 = g_2^2 = 0$, the system of two coupled quadratic equations for $g_1^2$ and $g_2^2$ allows three interacting fixed points.
Due to the homogeneity of the equations, two of them must be located on the $g_1$ and $g_2$ axes. These two \emph{decoupled} fixed points describe the chiral $\ONa$ and $\ONb$ universality classes, respectively. The coupling to a second critical scalar field, however, is a relevant perturbation, and the decoupled fixed points are therefore unstable when both $\phi$ and $\chi$ are tuned to criticality.
The topology of the flow then requires the third interacting fixed point to be stable. For symmetry reasons, it must be located on the bisectrix $g_1^2 = g_2^2$, and thus describes the chiral $\ONab$ universality class. This is the \emph{isotropic} fixed point.
It is generally expected, that the chiral $\ON$ universality classes exist for all $N$ and $\Nf$ compatible with Eq.~\eqref{eq:dgamma}, and the isotropic fixed point should therefore be located within the real coupling space $g_1^2 = g_2^2 > 0$.
Small symmetry-breaking perturbations are always irrelevant, and $\ONab$ symmetry becomes emergent when both $\phi$ and $\chi$ are tuned to criticality, at least within the Yukawa sector.
This general expectation is corroborated by the explicit evaluation of the one-loop flow: The isotropic fixed point is located at
\begin{align} \label{eq:fp}
	g_1^{*2} = g_2^{*2} = \frac{\epsilon}{2\Nf+4-N} + \mathcal O(\epsilon^2)\,,
\end{align}
and it is characterized by the stability exponents, which determine the flow near the fixed point,
\begin{align}
	(\theta_1, \theta_2) = \left(-1, -\frac{2\Nf+4}{2\Nf+4-N}\right) \epsilon + \mathcal O(\epsilon^2)\,,
\end{align}
with $N \coloneqq N_1 + N_2$.
Here, $\theta_1$ corresponds to the flow within the $\ONab$ invariant subspace, while $\theta_2$ corresponds to perturbations out of this subspace.
For all $N_1$, $N_2$, and $\Nf$ compatible with Eq.~\eqref{eq:dgamma}, we have $2\Nf +4 - N > 0$ and thus the isotropic fixed point is always real and stable within the Yukawa sector. We have explicitly checked that this remains true when the Nambu particle-hole construction is employed.\cite{Herbut:2011th}
The flow diagram in the $g_1$-$g_2$ sector is depicted for the example of the chiral $\mathrm{O}(3) \oplus \mathrm{O}(2)$ theory with emergent $\mathrm{O}(5)$ in Fig.~\ref{fig:flow}.
%
%
Using Eq.~\eqref{eq:fp}, we find the corresponding bosonic couplings at the isotropic fixed point as $\lambda_{1}^* = \lambda_{2}^* = \lambda_{3}^*=\lambda^*$ with
\begin{align}
	\lambda^* = \frac{f_N(\Nf) + 4 - N - 2\Nf}{8(N + 8)(2\Nf + 4 - N)}\epsilon + \mathcal O(\epsilon^2)\,,
\end{align}
where $f_N(\Nf)=2\Nf\sqrt{1 + \frac{5N+28}{\Nf} + \left(\frac{N-4}{2\Nf}\right)^2}>2\Nf + N$. Consequently, the effective potential is real and bounded from below.
The remaining stability exponents $(\theta_3,\theta_4,\theta_5)$, corresponding to the flow in the bosonic sector in the presence of $g_1^{*}$ and $g_2^{*}$, are smaller than $\theta_1$ and $\theta_2$ for all $N_1$, $N_2$, and $\Nf$ compatible with Eq.~\eqref{eq:dgamma}, see Ref.~\onlinecite{supp}.
The isotropic fixed point is therefore fully stable within the entire coupling space when both $r_1$ and $r_2$ are tuned to criticality. Consequently, the multicritical point between the Dirac semimetal and the gapped phases with $\ONa$ and $\ONb$ order parameters is characterized by an emergent $\ONab$ symmetry.
The scaling behavior near the multicritical point is described by the chiral $\ON$ universality class with $N=N_1 + N_2$.
The leading corrections to scaling correspond to a flow direction within the higher-symmetric subspace.
The corresponding exponent is $\omega_1 = -\theta_1$.
Small symmetry-breaking perturbations are strongly irrelevant and contribute to the corrections to scaling only at subleading order, with a comparatively large exponent $\omega_2 = -\theta_2$.
At the isotropic fixed point, we find the anomalous dimensions
\begin{align} \label{eq:anomdim}
	(\eta_\phi, \eta_\Psi) = \frac{1}{2\Nf+4-N}\left(2\Nf, \frac{N}{2}\right)\epsilon + \mathcal O(\epsilon^2)\,,
\end{align}
with $\eta_\phi = \eta_1 = \eta_2$.
The correlation-length exponent $\nu$ is obtained from the flow of the tuning parameters $r_1$ and $r_2$ at the isotropic fixed point and reads
\begin{align} \label{eq:nu}
	\nu & = \frac{1}{2} + \frac{2\Nf(N+14)+(N+2)\left[f_N(\Nf) + 4 - N \right]}{8(N + 8)(2\Nf + 4 - N)}\epsilon
	\nonumber \\ & \quad
	 + \mathcal O(\epsilon^2)\,.
\end{align}
Eqs.~\eqref{eq:anomdim} and \eqref{eq:nu} generalize previous results for the chiral Ising, XY, and Heisenberg universality classes\cite{Herbut:2009vu,Zerf:2017zqi} to the chiral $\ON$ universality classes with arbitrary $N \in \mathds N$.

\paragraph*{Critical exponents from functional RG.}

\begin{table}[t!]
\caption{\label{tab:FRG} Chiral $\ON$ universality classes in 2+1D from functional RG and $\epsilon$ expansion for $d_\gamma=4\Nf=8$: Correlation-length exponent $\nu$, anomalous dimensions $\eta_\phi$ and $\eta_\Psi$, and leading corrections-to-scaling exponent $\omega_1$. For the chiral Ising, XY, and Heisenberg universality classes, various further estimates are known, see Ref.~\onlinecite{Zerf:2017zqi}.}
\begin{tabular*}{\linewidth}{@{\extracolsep{\fill} } c c c c c c}
\hline\hline
$\Nf = 2$  & & $\nu$ & $\eta_\phi$ & $\eta_\Psi$ & $\omega_1$ \\
\hline
chiral Ising & FRG & 1.018 &   0.760 &   0.032 &   0.872\\ 
	& $\epsilon^1$ & 31/42 & 4/7 & 1/14 & 1\\ 
chiral XY & FRG & 1.160 & 0.875 & 0.062 & 0.878\\
	& $\epsilon^1$ & 4/5 & 2/3 & 1/6 &  1\\ 
chiral Heisenberg & FRG & 1.296 & 1.015 & 0.084 & 0.924 \\
	& $\epsilon^1$ & 97/110 & 4/5 & 3/10 & 1 \\ 
chiral $\mathrm{O}(4)$ & FRG & 1.364 & 1.159 & 0.091 & 1.017 \\
	& $\epsilon^1$ & 1 & 1 & 1/2 & 1\\ 
chiral $\mathrm{O}(5)$ & FRG & 1.356 & 1.285 & 0.089 & 1.132\\
	& $\epsilon^1$ & 31/26 & 4/3 & 5/6 & 1\\
\hline\hline
\end{tabular*}
\end{table}

Improved estimates for the critical exponents can be obtained from the nonperturbative functional RG (FRG) method.
In particular, this is a suitable approach for calculations directly in 2+1 dimensions and has been shown to compare well to other methods in the context of Gross-Neveu-type universality classes.\cite{Rosa:2000ju,Hofling:2002hj,Janssen:2012pq,mesterhazy2012,Janssen:2014gea,Classen:2015mar,Gehring:2015vja,Eichhorn:2016hdi,Knorr:2016sfs,Classen:2017hwp,Knorr:2017yze,yin2017}
The central FRG equation is formulated in terms of the effective average action $\Gamma_k$, which interpolates between the microscopic action $S$ at the UV cutoff, $k=\Lambda$, and the full quantum effective action $\Gamma$ in the IR, $k \to 0$.\cite{Wetterich:1992yh,Berges:2000ew}
For an approximate solution, we expand $\Gamma_k$ in powers of derivatives and truncate beyond the leading order,
\begin{align*}
	\Gamma_k=\int_ {\tau, \vec{x}}\left[Z_{\Psi}\bar\Psi\slashed{\partial}\Psi+g \bar\Psi\phi\Psi-\frac{Z_\phi}{2}\phi_a\partial_\mu^2\phi_a+U(\rho)\right].
\end{align*}
This constitutes the so-called local potential approximation (LPA').
Here, $\slashed{\partial} \equiv \gamma^\mu \partial_\mu$ is the standard Dirac operator and $\bar\Psi \equiv \Psi^\dagger \gamma^0$ denotes the Dirac conjugate. $\phi_a$, $a=1,\dots,N$, are the components of an O($N$) symmetric order-parameter field $\phi \equiv \sum_a^{N}\phi_a \beta_a^\phi$. It couples to the fermions via the scale-dependent Yukawa-type coupling $g$.
The scale-dependent effective potential $U(\rho)$ is a functional of all symmetry-allowed boson selfinteractions and as such depends only on the field invariant $\rho=\frac{1}{2}\phi_a\phi_a$.
Finally, we have introduced the wave-function renormalizations $Z_\Psi$ and $Z_\phi$, which also carry a scale dependence and are related to the anomalous dimensions via $\eta_\Phi=-(\partial_t Z_\Phi)/Z_\Phi$, with $\Phi \in \{\Psi,\phi\}$.
Explicit expressions of the FRG flow equations for the dimensionless versions of the scale-dependent quantities $U(\rho)$ and $g$ as well as for $\eta_\Psi$ and $\eta_\phi$ are given in Ref.~\onlinecite{supp}.
In practice, we expand $U(\rho)$ in a finite power series in $\rho$ around the origin up to order $\rho^6$. This yields a closed set of algebraic fixed-point equations, which we solve numerically.
For all $N$ and compatible $\Nf$ tested, we find a unique stable fixed point, characterizing the corresponding chiral $\ON$ universality class. For the example of $\Nf = 2$, relevant to spin-$1/2$ fermions on the honeycomb lattice, our estimates for the universal exponents are given in Tab.~\ref{tab:FRG}.
The estimates from FRG and the leading-order $\epsilon$ expansion are in reasonable agreement for $\nu$, $\eta_\phi$, and $\omega_1$. Significant differences appear in the fermion anomalous dimension $\eta_\Psi$.

\paragraph*{Implications for (3+1)D.}

In the limit of $\epsilon \to 0$, which applies to (3+1)D Weyl and Dirac systems, both the Yukawa couplings and the bosonic selfinteractions become marginally irrelevant. They flow to zero, however, with a fixed ratio $g_2/g_1 \to 1$ and $\lambda_{2,3}/\lambda_1 \to 1$, indicating emergent $\ON$ symmetry also in this case.

\paragraph*{Conclusions.}

We have demonstrated that the multicritical Dirac systems with compatible $\ONa$ and $\ONb$ order parameters are generically characterized by emergent $\ONab$ symmetry. Within the first-order $\epsilon$ expansion, this result holds for all $N_1$ and $N_2$ and fermion flavor numbers $\Nf$, as long as the corresponding representation of the Clifford algebra exists. Put differently, the chiral $\ON$ universality classes are stable under any small perturbation that breaks $\ON$ symmetry.
This conclusion is in surprising contrast to the purely bosonic $\ON$ universality classes, in which symmetry-breaking perturbations destabilize the isotropic $\ON$ fixed point when $N \geq 3$.\cite{Pelissetto:2000ek, Carmona:1999rm, Calabrese:2002bm, herbut2007, Eichhorn:2013zza}
There, however, it is well known that the first-order $\epsilon$ expansion, when extrapolated to $\epsilon = 1$, significantly overestimates the stability of the isotropic fixed point. In the present system, by contrast, higher-loop corrections become suppressed for large number of Dirac fermions $\Nf$. Consequently, we expect our conclusion of emergent $\ON$ symmetry for all $N>1$ to be true also in $d=2$ as long as $\Nf$ is large enough.
Within the first-order $\epsilon$ expansion, symmetry-breaking perturbations are strongly irrelevant also for small $\Nf$, and one is therefore tempted to argue that higher-order corrections will ultimately not overturn the leading-order result also in this case. This conjecture deserves further investigation.
The chiral $\ON$ universality classes are therefore in principle accessible in lattice Dirac systems without an explicit $\ON$ symmetry by tuning two parameters through a suitable multicritical point. The critical behavior is characterized by universal exponents, for which we have given estimates from $\epsilon$ expansion and functional RG.
Our result partly explains the recently observed emergent $\mathrm{O}(4)$ symmetry in simulations of a Dirac system with anticommuting mass terms.\cite{Sato:2017tgx}
There, however, evidence for emergent $\mathrm{O}(4)$ is found also significantly away from the multicritical point. This represents an interesting problem on its own.
%

\paragraph*{Note added.}
After this work was submitted, a related preprint appeared on the arXiv,\cite{Roy:2017vkg} also demonstrating emergent $\ON$ symmetry in multicritical Dirac systems.

\paragraph*{Acknowledgments.}
The authors are grateful to Fakher Assaad, Laura Classen, Martin Hohenadler, Bitan Roy, and Toshihiro Sato for discussions. 
LJ acknowledges support by the DFG within project A04 of SFB1143.
IFH was supported by the NSERC of Canada.
MMS was supported by the DFG through the Collaborative
Research Center SFB1238, TP C04.

%


\begin{thebibliography}{99}%

\bibitem{Zhang1997}
  S.-C.~Zhang, Science {\bf 275}, 1089 (1997).

\bibitem{Nahum:2015vka}
  A.~Nahum, P.~Serna, J.~T.~Chalker, M.~Ortu\~no, and A.~M.~Somoza,
  Phys.\ Rev.\ Lett.\  {\bf 115}, 267203 (2015).

\bibitem{Wang:2017txt}
  C.~Wang, A.~Nahum, M.~A.~Metlitski, C.~Xu, and T.~Senthil,
  Phys.\ Rev.\ X {\bf 7}, 031051 (2017).

\bibitem{Janssen:2017eeu}
  L.~Janssen and Y.~C.~He,
  Phys.\ Rev.\ B {\bf 96}, 205113 (2017).

\bibitem{Sato:2017tgx}
  T.~Sato, M.~Hohenadler, and F.~F.~Assaad,
  Phys.\ Rev.\ Lett.\  {\bf 119}, 197203 (2017).

\bibitem{Pelissetto:2000ek}
  A.~Pelissetto and E.~Vicari,
  Phys.\ Rept.\  {\bf 368}, 549 (2002).

\bibitem{herbut2007}
I.~Herbut, {\it A Modern Approach to Critical Phenomena} (Cambridge University Press, Cambridge, England, 2007).

\bibitem{Eichhorn:2013zza}
  A.~Eichhorn, D.~Mesterh\'azy, and M.~M.~Scherer,
  Phys.\ Rev.\ E {\bf 88}, 042141 (2013).

\bibitem{Carmona:1999rm}
  J.~M.~Carmona, A.~Pelissetto, and E.~Vicari,
  Phys.\ Rev.\ B {\bf 61}, 15136 (2000).

\bibitem{Calabrese:2002bm}
  P.~Calabrese, A.~Pelissetto, and E.~Vicari,
  Phys.\ Rev.\ B {\bf 67}, 054505 (2003).

\bibitem{li2015}
Z.-X. Li, Y.-F. Jiang, S.-K. Jian, and H. Yao, Nature Comm.\ {\bf 8}, 314 (2017).

\bibitem{Scherer:2016zwz}
  M.~M.~Scherer and I.~F.~Herbut,
  Phys.\ Rev.\ B {\bf 94}, 205136 (2016).

\bibitem{Ponte:2012ru}
  P.~Ponte and S.~S.~Lee,
  New J.\ Phys.\  {\bf 16}, 013044 (2014).

\bibitem{Grover:2013rc}
  T.~Grover, D.~N.~Sheng, and A.~Vishwanath,
  Science {\bf 344}, 280 (2014).

\bibitem{Li:2017dkj}
  Z.~X.~Li, A.~Vaezi, C.~B.~Mendl, and H.~Yao,
  arXiv:1711.04772 [cond-mat.str-el].
  
\bibitem{ryu2009}
  S.~Ryu, C.~Mudry, C.-Y.~Hou, and C.~Chamon,
  Phys. Rev. B {\bf 80}, 205319 (2009).

\bibitem{das2014}
  A.~Das and S.~Okubo,  {\it Lie Groups and Lie Algebras for Physicists} (World Scientific, Singapore, 2014).

\bibitem{Herbut:2011th}
  I.~F.~Herbut,
  Phys.\ Rev.\ B {\bf 85}, 085304 (2012).

\bibitem{ZinnJustin:1991yn}
  J.~Zinn-Justin,
  Nucl.\ Phys.\ B {\bf 367}, 105 (1991).

\bibitem{Roy:2011pg}
  B.~Roy,
  Phys.\ Rev.\ B {\bf 84}, 113404 (2011).

\bibitem{Roy:2013aya}
  B.~Roy and V.~Juri\v{c}i\'{c},
  Phys.\ Rev.\ B {\bf 90}, 041413 (2014).

\bibitem{Classen:2015ssa}
  L.~Classen, I.~F.~Herbut, L.~Janssen, and M.~M.~Scherer,
  Phys.\ Rev.\ B {\bf 92},  035429 (2015).

\bibitem{Classen:2015mar}
  L.~Classen, I.~F.~Herbut, L.~Janssen, and M.~M.~Scherer,
  Phys.\ Rev.\ B {\bf 93}, 125119 (2016).

\bibitem{Herbut:2009vu}
  I.~F.~Herbut, V.~Juri\v{c}i\'{c}, and O.~Vafek,
  Phys.\ Rev.\ B {\bf 80}, 075432 (2009).

\bibitem{footnote0}
We note that the Kekul\'e transition on the honeycomb lattice is naturally described by a complex order parameter with a discrete $\mathds{Z}_3$ symmetry which, however, is enhanced to an emergent U(1) at the QCP.\cite{li2015,Scherer:2016zwz}

\bibitem{footnote1}
The bosonic contributions $\propto \lambda_1 \lambda_3$ and $\propto \lambda_2 \lambda_3$ in Eq.~\eqref{eq:lambda12} do not agree with the corresponding equation given in Ref.~\onlinecite{Roy:2011pg}.
Since only with these contributions our flow equations reproduce the known purely bosonic limit,\cite{Pelissetto:2000ek,Calabrese:2002bm,herbut2007} we expect them to be correct.

\bibitem{Mihaila:2017ble}
  L.~N.~Mihaila, N.~Zerf, B.~Ihrig, I.~F.~Herbut, and M.~M.~Scherer,
  Phys.\ Rev.\ B {\bf 96}, 165133 (2017).

\bibitem{Zerf:2017zqi}
  N.~Zerf, L.~N.~Mihaila, P.~Marquard, I.~F.~Herbut, and M.~M.~Scherer,
  Phys.\ Rev.\ D {\bf 96}, 096010 (2017).
  
\bibitem{supp}
See Supplemental Material for explicit expressions of the subleading exponents and functional RG flow equations.

\bibitem{Rosa:2000ju}
L.~Rosa, P.~Vitale, and C.~Wetterich,
Phys.\ Rev.\ Lett.\  {\bf 86}, 958 (2001).

\bibitem{Hofling:2002hj}
F.~H\"ofling, C.~Nowak, and C.~Wetterich,
Phys.\ Rev.\  B {\bf 66}, 205111 (2002).

\bibitem{Janssen:2012pq}
  L.~Janssen and H.~Gies,
  Phys.\ Rev.\ D {\bf 86}, 105007 (2012).

\bibitem{mesterhazy2012}
D.~Mesterh\'azy, J.~Berges, and L.~von~Smekal, Phys.\ Rev.\ B {\bf 86}, 245431 (2012).

\bibitem{Janssen:2014gea}
  L.~Janssen and I.~F.~Herbut,
  Phys.\ Rev.\ B {\bf 89}, 205403 (2014).

\bibitem{Gehring:2015vja}
  F.~Gehring, H.~Gies, and L.~Janssen,
  Phys.\ Rev.\ D {\bf 92}, 085046 (2015).

\bibitem{Eichhorn:2016hdi}
  A.~Eichhorn, L.~Janssen, and M.~M.~Scherer,
  Phys.\ Rev.\ D {\bf 93}, 125021 (2016).

\bibitem{Knorr:2016sfs}
  B.~Knorr,
  Phys.\ Rev.\ B {\bf 94}, 245102 (2016).

\bibitem{Classen:2017hwp}
  L.~Classen, I.~F.~Herbut, and M.~M.~Scherer,
  Phys.\ Rev.\ B {\bf 96}, 115132 (2017).

\bibitem{Knorr:2017yze}
  B.~Knorr,
  arXiv:1708.06200 [cond-mat.str-el].
  
\bibitem{yin2017}
  S.~Yin, S.-K.~Jian, and H.~Yao, 
  arXiv:1711.10473 [cond-mat.str-el].

%

\bibitem{Wetterich:1992yh}
  C.~Wetterich,
  Phys.\ Lett.\ B {\bf 301}, 90 (1993).

\bibitem{Berges:2000ew}
  J.~Berges, N.~Tetradis, and C.~Wetterich,
  Phys.\ Rept.\  {\bf 363}, 223 (2002).
  
\bibitem{Roy:2017vkg} 
  B.~Roy, P.~Goswami and V.~Juri\v{c}i\'{c},
  arXiv:1712.05400 [cond-mat.str-el].

\end{thebibliography}
\end{document}



\title{Supplemental Material:\\
Compatible orders and fermion-induced emergent symmetry in Dirac systems}

%
\author{Lukas Janssen}
\affiliation{Institut f\"ur Theoretische Physik, Technische Universit\"at Dresden, 01062 Dresden, Germany}
\author{Igor~F.~Herbut}
\affiliation{Department of Physics, Simon Fraser University, Burnaby, British Columbia, Canada V5A 1S6}
\author{Michael~M.~Scherer}
\affiliation{Institut f\"ur Theoretische Physik, Universit\"at K\"oln, 50937 K\"oln, Germany}

\maketitle

\section{Stability exponents in $\epsilon$ expansion} \label{app:exponents}

The exponents $\theta_1$ and $\theta_2$ at the isotropic fixed point, obtained from the flow in the Yukawa-coupling sector, are given in the main text. Here, we demonstrate that the remaining exponents $\theta_3$, $\theta_4$, and $\theta_5$, determining the stability of the isotropic fixed point in the bosonic sector, indeed correspond to irrelevant directions. They are given by
\begin{align}
	\theta_3&=-\frac{(2 + N) (2 \Nf+N-4) + 6 f_N(\Nf)}{(N + 8) (2\Nf + 4 - N)}  \epsilon + \mathcal O(\epsilon^2)\,,\\
	%
	%
	\theta_4&= -\frac{N (2 \Nf+N - 4) + (16 + N) f_N(\Nf)}{2 (N + 8) (2\Nf + 4 - N)} \epsilon + \mathcal O(\epsilon^2)\,,\\
	%
	%
	\theta_5&=-\frac{f_N(\Nf)}{2\Nf+4-N}\epsilon + \mathcal O(\epsilon^2)\,.
\end{align}
The function $f_N(\Nf)$ is given in the main text.
A simple analysis shows that $\theta_3$, $\theta_4$, and $\theta_5$ are always negative and smaller than $\theta_1$ and $\theta_2$ for all $N_1$, $N_2$, and $\Nf$ compatible with Eq.~\eqdgamma\ in the main text.
Consequently, they correspond to directions that are even more irrelevant than the flow in the Yukawa sector, and the isotropic chiral $\ONab$ fixed point is indeed stable.

\section{Functional RG flow equations}
\label{app:thresholds}

The FRG generalizes Wilson's momentum-shell RG by adding a scale-dependent mass-like regulator term to the action, i.e., 
\begin{align}
	S\mapsto S_k \equiv S + \int_p\frac{1}{2}\Phi(-p)R_{k}(p)\Phi(p)\,.
\end{align}
Such modified action then leads to a scale-dependent partition function $\mathcal Z_k=\int_\Lambda \mathcal{D}\Phi\, \exp(-S_k[\Phi])$, where $\Phi$ collects all field variables of a specific model.
The central FRG equation is formulated in terms of the Legendre transform of $\ln \mathcal Z_k$, the effective average action $\Gamma_k$. For suitably chosen regulators $R_k$, it interpolates between the microscopic action $S$ at the UV cutoff, $k = \Lambda$, and the full quantum effective action $\Gamma$ in the IR, $k \to 0$. Its flow is governed by\cite{Wetterich:1992yh,Berges:2000ew}
\begin{align}\label{eqn:Wetterich}
\partial_k\Gamma_k = \frac{1}{2}\text{STr}\left[(\Gamma_k^{(2)}+R_k)^{-1}\partial_k R_k\right],
\end{align}
where $\Gamma_k^{(2)}$ denotes the functional Hessian of $\Gamma_k$ with respect to the fields $\Phi$.
To determine the fixed-point properties within the FRG approach in $D=2+1$ space-time dimensions, we introduce dimensionless quantities. The dimensionless effective potential $u$ and the dimensionless Yukawa-type coupling $\hat g$ are given by $u(\hat\rho)=k^{-D}U(\rho)$ and $\hat g^2=k^{d-3}g^2/(Z_{\phi}Z_{\Psi}^2)$,
where $\rho=k^{D-2}Z_{\phi}^{-1} \hat\rho$.
The flow equations for $u$ and $\hat g^2$, as well as the anomalous dimensions $\eta_\phi$ and $\eta_\Psi$ are obtained by substituting our ansatz for $\Gamma_k$ into the flow equation~\eqref{eqn:Wetterich} and applying suitable projection prescriptions.
Therewith, we obtain coupled differential equations for the effective potential,
\begin{align}
\partial_t u&=-D u +(D-2+\eta_\phi)\hat\rho u^{\prime}-4d_\gamma v_D l^\mathrm F(2 \hat g^2 \hat\rho)\notag\\[5pt]
&\quad+2v_D \left[l^\mathrm{B}( u^{\prime}+2\hat\rho  u^{\prime\prime})+(N-1)l^\mathrm B( u^{\prime})\right]\,,
\end{align}
and the Yukawa-type coupling,
\begin{align}
\partial_t \hat g^2=&(D-4+\eta_\phi+2\eta_\Psi)\hat g^2\notag\\
&-8(N-2)v_{D}\hat g^4l_{11}^\mathrm{FB}(2\hat g^2\hat\rho,u^{\prime})\,.
\end{align}
Here, we have abbreviated $v_D \coloneqq  1/[2^{D+1}\pi^{D/2}\Gamma(D/2)]$ and $\partial_t \coloneqq k \partial_k$. Primes denote derivatives with respect to $\hat\rho$.
Further, we obtain two algebraic equations for the anomalous dimensions, $\eta_\phi=\frac{16v_D\hat g^2}{D}d_\gamma m_{4}^\mathrm F(2\hat g^2\hat\rho)$ and $\eta_\Psi=\frac{8v_D\hat g^2}{D}N  m_{12}^\mathrm F(2\hat g^2\rho,u^\prime)$.
The above results agree with the known FRG equations for $N=1,2,3$.\cite{Hofling:2002hj,Classen:2017hwp,Janssen:2014gea}
The threshold functions $l^\mathrm B$, $l^\mathrm F$, $m_4^\mathrm F$, $m_{12}^\mathrm F$, and $l_{11}^\mathrm{FB}$ depend on the regulator.
Here, we choose linear cutoff functions $R_\phi=Z_\phi p^2r_\phi$ and $R_\Psi=Z_\Psi \slashed{p} r_\Psi$  for fermions and bosons, respectively, of the form
$r_{\Psi}(q)=\big(\frac{k}{q}-1\big)\,\Theta(k^2-q^2)$ and \mbox{$r_{\phi}(q)=\big(\frac{k^2}{q^2}-1\big)\,\Theta(k^2-q^2)$}.
$\Theta(x)$ denotes the step function.
For this case, the threshold functions appearing in the flow of the effective potential read\cite{Berges:2000ew}
\begin{align*}
l^{\mathrm B}(\omega)&=\frac{2}{D}\left(1-\frac{\eta_\phi}{D+2}\right)\frac{1}{1+\omega}\,,\\
l^{\mathrm F}(\omega)&=\frac{2}{D}\left(1-\frac{\eta_\Psi}{D+1}\right)\frac{1}{1+\omega}\,.
\end{align*}
The threshold function for the Yukawa-type coupling involves both boson and fermion mass contributions,
\begin{multline*}
l_{11}^\mathrm{FB}(\omega_\Psi,\omega_\phi) =
\frac{2}{D}\bigg[\left(1-\frac{\eta_\Psi}{D+1}\right)\frac{1}{1+\omega_\Psi} \\
+\left(1-\frac{\eta_\phi}{D+2}\right)\frac{1}{1+\omega_\phi}\bigg]\frac{1}{(1+\omega_\Psi)(1+\omega_\phi)}\,.
\end{multline*}
For the anomalous dimensions, the corresponding functions read
\begin{align*}
m_4^{\mathrm F}(\omega)&=\frac{1}{(1+\omega)^4}+\frac{1-\eta_\Psi}{D-2}\frac{1}{(1+\omega)^3}\notag\\
&\quad-\left(\frac{1-\eta_\Psi}{2D-4}+\frac{1}{4}\right)\frac{1}{(1+\omega)^2}\,,\displaybreak[0]\\
m_{12}^{\mathrm{FB}}(\omega_\Psi,\omega_\phi)&= \left(1-\frac{\eta_\phi}{D+1}\right) \frac{1}{(1+\omega_\Psi)(1+\omega_\phi)^2}.
\end{align*}
To simplify the numerics, we expand the potential $u$ in powers of $\hat\rho$ around the origin,
$
u(\hat\rho)=\sum_{i=1}^{i_\text{max}}\frac{\hat\lambda_{i}}{i!} \hat\rho^i.
$
This defines the couplings $\hat\lambda_i$.
Therewith, $\hat\lambda_1$ corresponds to the tuning parameter $r$ and $\hat\lambda_2$ to the quartic coupling.
The flow equations for the $\hat\lambda_{i}$'s are obtained by suitable projections of the flow of the effective potential, evaluated at its minimum $\hat\rho = 0$,
%
$
	\partial_t \hat\lambda_{i} = \big(\frac{\partial^{i}}{\partial \hat\rho^i}\partial_t u(\hat\rho)\big)\big\vert_{\hat\rho=0}\,.
$
%
Similarly, $\partial_t \hat g^2$, $\eta_\phi$, and $\eta_\Psi$ are evaluated at the potential's minimum $\hat\rho=0$.
For the explicit values quoted in the main text, we have chosen $i_\mathrm{max}=6$.
We have confirmed that this leads to convergent fixed-point properties within the polynomial expansion of the potential.

%